\documentclass[twocolumn,aps,prc,superscriptaddress,floatfix,nofootinbib,linenumbers,10pt]{revtex4}
\usepackage{epsfig,graphics}
\usepackage{appendix}
\usepackage{graphicx}
\usepackage{dcolumn,multirow}
\usepackage{bm}
\usepackage{braket,overpic}
\usepackage{amsmath,amssymb}
\usepackage[usenames]{color}
\usepackage{pgffor}  
\usepackage{ulem} 

\usepackage{placeins}

\newcommand{\be}{\begin{eqnarray}}
\newcommand{\ee}{\end{eqnarray}}
\newcommand{\bea}{\begin{eqnarray}}
\newcommand{\eea}{\end{eqnarray}}
\newcommand{\bc}{\begin{center}}
\newcommand{\ec}{\end{center}}

\newcommand{\beq}{\begin{equation}}
\newcommand{\eeq}{\end{equation}}
\newcommand{\ba}{\begin{eqnarray}}
\newcommand{\ea}{\nonumber \end{eqnarray}}

\newcommand{\bi}{\begin{enumerate}}
\newcommand{\ei}{\end{enumerate}}

\definecolor{darkyellow}{rgb}{1.0,0.5,0}
\definecolor{lightyellow}{cmyk}{0,0,0.5,0}
\definecolor{lightred}{rgb}{1,0.5,0.5}
\definecolor{lightgreen}{rgb}{0,0.4,0}
\definecolor{lightblue}{rgb}{0.5,0.5,1}
\definecolor{darkred}{rgb}{0.8,0,0}
\definecolor{darkgreen}{rgb}{0,0.4,0}
\definecolor{darkcyan}{cmyk}{1,0.3,0.3,0.3}
\definecolor{darkblue}{rgb}{0,0,0.6}
\definecolor{lightbrown}{rgb}{0.7,0.3,0.3}
\definecolor{darkbrown}{rgb}{0.5,0,0}
\definecolor{violett}{rgb}{0.6,0,0.8}

\begin{document}

\title{Partial wave analysis of reactions with four meson final states}
\newcommand*{\PNPI}{NRC "Kurchatov Institute", Petersburg Nuclear Physica Institute, 188300 Gatchina, Russia}

\affiliation{\PNPI}

\author{M.A.~Matveev} 
\author{A.T.~Sitnikov} 
\author{A.V.~Sarantsev} 
\altaffiliation{Corresponding author: \texttt{sarantsev\_AV@pnpi.nrcki.ru}}

\date{\today}

\begin{abstract}
We construct a formalism which describes the resonances decaying into four pseudoscalar meson final states. This method is fully covariant and can be directly applied for the partial-wave analysis of high statistical data. Two topologies of the process are considered: two intermediate resonances each decaying into two final mesons and cascade decay via three meson intermediate states. In particular, we consider the production of such states in the central collision reactions and in radiative $J/\Psi$ decay.
\end{abstract}

\pacs{25.75.-q}
\maketitle

The main information about meson states comes from the $\pi N$ scattering with high energy pion beams, from the antiproton-nucleon annihilation, from the decay of relatively stable, heavy hadrons and from the production of mesons in the central collision. In most of the reactions analyzed, resonances decay into final states with two stable particles (see, for example, \cite{GAMS:1998hws,Grayer:1974cr,Longacre:1986fh,BESIII:2015rug,BESIII:2018ubj}), or the reactions can be considered as quasi-two-particle scattering (for example, antiproton-proton annihilation at rest into three pseudoscalar mesons \cite{Bugg:1994mg,Amsler:1995gf,CrystalBarrel:1995fiy}). We note that some of the developments for multichannel decays were made for the analysis of radiative decay $J/\Psi$ \cite{BES:1999dmf} and for the annihilation of the antiproton-proton at rest into five pions \cite{Thoma:1997si}. However, no systematic formalism for the analysis of the data in which meson resonances decay to a four-meson final state had been developed. Nevertheless, such a decay mode is a dominant one for the resonances in many partial waves already at masses around 1400 MeV.  

In the scalar isoscalar sector, the decay mode $4\pi$ is already a dominant one for the $f_0(1370)$ state. Moreover, in the elastic scattering data $\pi \pi$ (extracted from the reaction $\pi N$) this state can only appear due to rescattering with other scalar states in the $4\pi$ channel and cannot be clearly observed. This property simulated a number of discussions about the existence of this state. However, this state was observed in the antiproton-proton in the analysis of the data with the final states $3\pi^0$ \cite{Bugg:1994mg} and $5\pi$ \cite{Thoma:1997si}. We mention that the scalar isoscalar sector is rather difficult to analyze.  In this sector, one expects a strong mixing between non-strange and strange quark-antiquark components and a production of exotic states. For example, many authors treat lowest-scalar states as molecular-like states or as four-quark bound systems. The lowest bound states of two gluons: glueballs are also expected in this sector \cite{Szczepaniak:2003mr}. Such a picture makes the analysis of this sector a rather complicated procedure, and information about four-meson final states can be vital for understanding the spectrum and properties of these mesons.

In the present paper, we construct the covariant approach for the analysis of the resonances decaying into four pseudoscalar mesons and consider the production of these resonances in $J/\Psi$ radiative decay and in the $NN$ central collision reactions (pomeron-pomeron scattering). The $J/\Psi$ radiative decay is one of the main sources of the search for glueball states. In fact, the partial wave decomposition of the BES III data on the $J/\psi$ radiative decay into two pseudo-scalar mesons \cite{BESIII:2015rug,BESIII:2018ubj} demonstrated a very complicated resonant structures in the isoscalar-scalar partial wave in the mass region 1500-2100 MeV. The combined analysis of these data with the scattering data from $\pi \pi$ and the data on the annihilation of the proton-antiproton at rest in three pseudoscalar mesons revealed the contribution of ten scalar states \cite{Sarantsev:2021ein}. The distribution of resonance production intensities demonstrated a peak in the mass region of 1850 MeV, which was explained by the presence of two-gluon components in the observed scalar states. This idea was confirmed by the calculation of the mixing angles between $n\bar n$, $s\bar s$ and the glueball components \cite{Klempt:2021wpg}. In the alternative analysis of these data \cite{Rodas:2021tyb} it was declared that the resonance with mass around 1750 MeV is produced dominantly and is the main candidate to the scalar glueball. The analysis of the data with the four-pion channel should resolve this issue. 

Another issue is the search for the tensor glueball. If the $J/\Psi$ radiative decay into two pseudoscalar mesons reveal the production of the scalar glueball in the mass region 1700-1900 MeV then one should expect the production of the tensor glueball (for example from lattice calculations \cite{Athenodorou:2020ani,Gregory:2012hu}) in the mass region 2200-2500 MeV.The tensor partial waves extracted from the BES III data showed a strong production of $f_2(1275)$ (decaying into $2\pi$ final state) and $f'_2(1525)$ (decaying into $K\bar K$ final state) and practically no structure at higher energies. In fact, no clear signal was found in the analysis \cite{Klempt:2022qjf}. The only solution to this problem is related to the fact that tensor states in this mass region decay dominantly into four pseudo-scalar mesons, and signals from the tensor glueball should be searched in the data on the $J/\Psi$ radiative decay into these final states. 

Another prominent source for the production of the glueball states is in the meson production at nucleon-nucleon central collision reactions. In these processes, the states are predominantly produced from the pomeron-pomeron collision. Taking into account that the pomeron is an effective way to describe the gluon lattice, it is relevant to expect that the states with a large gluon component will be strongly produced in such a reaction. Therefore, our method should be useful for planning new experiments for the new constructed colliders like the NICA.

\section{The covariant spin-orbital formalism}
\subsection{Orbital angular momentum tensor}

The formalism for the construction of the orbital momentum tensor is given in detail in \cite{Anisovich:2004zz}. We briefly recall it here. Consider the decay of a composite system with
spin $J$ and momentum $P$ ($P^2=s$) into two spinless particles with momenta $k_1$ and $k_2$. The only quantities measured in such a reaction
are the particle momenta. The angular dependent part of
the wave function of the composite state is described by tensors
constructed out of these momenta and the metric tensor. Such tensors
(we denote them as $X^{(L)}_{\mu_1\ldots\mu_L}$, where $L$ is the
orbital momentum)
are called orbital angular momentum tensors and
correspond to irreducible representations of the Lorentz group.
They satisfy the following properties:
\begin{itemize}
\item Symmetry under permutation of any two indices:
\begin{eqnarray}
 X^{(L)}_{\mu_1\ldots\mu_i\ldots\mu_j\ldots\mu_L}\; =\;
X^{(L)}_{\mu_1\ldots\mu_j\ldots\mu_i\ldots\mu_L}.
\label{oth_b1}
\end{eqnarray}
\item Orthogonality to the total momentum of the system, $P=k_1+k_2$.
\be
P_{\mu_i}X^{(L)}_{\mu_1\ldots\mu_i\ldots\mu_L}\ =\ 0
\label{oth_b2}
\ee
\item The traceless property for convolution of any  two  indices with metric tensor:
\be
g_{\mu_i\mu_j}X^{(L)} _{\mu_1\ldots\mu_i\ldots\mu_j\ldots\mu_L}\
\ =\ 0.
\label{oth_b3}
\ee
\end{itemize}

The orthogonality condition (\ref{oth_b2})
is automatically fulfilled if the tensors are
constructed from the relative momenta
$k^\perp_\mu$ and the tensor $g^\perp_{\mu\nu}$ orthogonal to the total momentum of the system:
\begin{eqnarray}
    k^\perp_\mu=\frac12
(k_1-k_2)_\nu g^{\perp}_{\mu\nu}\qquad\qquad
g^{\perp}_{\mu\nu}=g_{\mu\nu}-\frac{P_\mu P_\nu}{s}
\end{eqnarray}

In the center-of-mass system (cms), where $P=(P_0,\vec P)=(\sqrt s,0)$, the vector $k^\perp$ is space-like: $k^\perp=(0,\vec k)$.

The orbital tensor for $L=0$ is a scalar value (for example, a unit),
and the tensor for the orbital momentum $L=1$ is a vector that can only be
constructed from $k^\perp_\mu$.
The orbital angular momentum tensors for $L$ up to three have the form: 
\begin{eqnarray}
&&X^{(0)}=1\ , \qquad X^{(1)}_\mu=k^\perp_\mu\ , \qquad\nonumber \\
&&X^{(2)}_{\mu_1 \mu_2}=\frac32\left(k^\perp_{\mu_1}
k^\perp_{\mu_2}-\frac13\, k^2_\perp g^\perp_{\mu_1\mu_2}\right), \nonumber  \\
&&X^{(3)}_{\mu_1\mu_2\mu_3}= 
\frac52\Big[k^\perp_{\mu_1} k^\perp_{\mu_2 }
k^\perp_{\mu_3} -\nonumber \\
&&\frac{k^2_\perp}5\left(g^\perp_{\mu_1\mu_2}k^\perp
_{\mu_3}+g^\perp_{\mu_1\mu_3}k^\perp_{\mu_2}+
g^\perp_{\mu_2\mu_3}k^\perp_{\mu_1}
\right)\Big] .
\end{eqnarray}
The tensors $X^{(L)}_{\mu_1\ldots\mu_L}$ for $L\ge 1$ can be constructed from the tensors with lower orbital momenta in the 
form of a recurrent expression:
\begin{eqnarray}
&&X^{(L)}_{\mu_1\ldots\mu_L}=k^\perp_\alpha
Z^{\alpha}_{\mu_1\ldots\mu_L} \; ,
\nonumber\\
&&Z^{\alpha}_{\mu_1\ldots\mu_L}=
\frac{2L-1}{L^2}\Big (
\sum\limits^L_{i=1}X^{{(L-1)}}_{\mu_1\ldots\mu_{i-1}\mu_{i+1}\ldots\mu_L}
g^\perp_{\mu_i\alpha}-
\nonumber \\
 &&\frac{2}{2L-1}  \sum^L_{i,j=1 \atop i<j}
g^\perp_{\mu_i\mu_j}
X^{{(L-1)}}_{\mu_1\ldots\mu_{i-1}\mu_{i+1}\ldots\mu_{j-1}\mu_{j+1}
\ldots\mu_L\alpha} \Big ).~~~
\label{z}
\end{eqnarray}
The normalization of the tensors is fixed by the convolution equality:
\begin{eqnarray}
X^{(L)}_{\mu_1\ldots\mu_L}k^\perp_{\mu_L}=k^2_\perp
X^{(L-1)}_{\mu_1\ldots\mu_{L-1}}.
\label{ceq}
\end{eqnarray}
Iterating eq. (\ref{z}) one obtains the
following expression for the tensor $
X^{(L)}_{\mu_1\ldots\mu_L}$:
\begin{eqnarray}
X^{(L)}_{\mu_1\ldots\mu_L}(k^\perp)
=\alpha(L)
\bigg [
k^\perp_{\mu_1}k^\perp_{\mu_2}k^\perp_{\mu_3}k^\perp_{\mu_4}
\ldots k^\perp_{\mu_L}
&-&  \nonumber \\
\frac{k^2_\perp}{2L-1}\bigg(
g^\perp_{\mu_1\mu_2}k^\perp_{\mu_3}k^\perp_{\mu_4}\ldots
k^\perp_{\mu_L} &+& \nonumber \\
g^\perp_{\mu_1\mu_3}k^\perp_{\mu_2}k^\perp_{\mu_4}\ldots
k^\perp_{\mu_L} + \ldots \bigg)
&+& \\
\frac{k^4_\perp}{(2L\!-\!1)
(2L\!-\!3)}\bigg(
g^\perp_{\mu_1\mu_2}g^\perp_{\mu_3\mu_4}k^\perp_{\mu_5}
k^\perp_{\mu_6}\ldots k_{\mu_L} &+&
\nonumber \\
g^\perp_{\mu_1\mu_2}g^\perp_{\mu_3\mu_5}k^\perp_{\mu_4}
k^\perp_{\mu_6}\ldots k_{\mu_L}+
\ldots\bigg)&+&\ldots\bigg ].  \nonumber
\label{x-direct}
\end{eqnarray}
where
\begin{eqnarray}
    \alpha(L)=\prod^L_{l=1}\frac{2l-1}{l}=\frac{(2L-1)!!}{L!} \; .
\label{alpha}
\end{eqnarray}
Using normalization condition we obtain:
\begin{eqnarray}
X^{(L)}_{\mu_1\ldots\mu_L}X^{(L)}_{\mu_1\ldots\mu_L}=
\alpha(L)(k^2_\perp)^L
\end{eqnarray}

\subsection{The boson projection operator}

The projection operator $O^{\mu_1\ldots\mu_L}_{\nu_1\ldots \nu_L}(P)$ for the partial wave with angular momentum $L$ is defined as:
\begin{eqnarray}
\int\frac{d\Omega }{4\pi}\
&&X^{(L)}_{\mu_1\ldots\mu_L}(k^\perp)
X^{(L)}_{\nu_1\ldots\nu_L}(k^\perp)  =  \nonumber \\
 &&\frac{\alpha(L)}{2L+1}  (k^2_\perp)^L 
O^{\mu_1\ldots\mu_L}_{\nu_1\ldots \nu_L}(P).
\label{18}
\end{eqnarray}
This tensor satisfies the following relations:
\begin{eqnarray}
X^{(L)}_{\mu_1\ldots\mu_L}(k^\perp)
O^{\mu_1\ldots\mu_L}_{\nu_1\ldots \nu_L}(P)
&=&\ X^{(L)}_{\nu_1\ldots \nu_L}(k^\perp)\ , \nonumber \\
O^{\mu_1\ldots\mu_L}_{\alpha_1\ldots\alpha_L}(P) \
O^{\alpha_1\ldots\alpha_L}_{\nu_1\ldots \nu_L}(P)\
&=&\ O^{\mu_1\ldots\mu_L}_{\nu_1\ldots \nu_L}(P)\ .
\label{proj_op}
\end{eqnarray}
This tensor has the same symmetry, orthogonality and traceless
properties as $X$-tensors (for the same set
of up and down indices) but the $O$-operator
 does not depend on the relative momentum
of the constituents and does not describe decay
processes. It represents the structure of the propagator of the composite system. 
Taking into account the definition of the projection operators
(\ref{proj_op})
and the properties of the $X$-tensors (\ref{x-direct}) we obtain:
\begin{eqnarray}
k_{\mu_1}\ldots k_{\mu_L}
O^{\mu_1\ldots\mu_L}_{\nu_1\ldots \nu_L}\ =
\frac{1}{\alpha(L)}
X^{(L)}_{\nu_1\ldots\nu_L}(k^\perp).
\label{19}
\end{eqnarray}
This equation presents the basic property of
the projection operator:
it projects any tensor with $L$ indices onto the tensor which satisfied all properties of the partial
wave considered. \\
The projection operator can also be calculated from tensors with lower rank using the recurrent expression:
\begin{eqnarray}
&&O^{\mu_1\ldots\mu_L}_{\nu_1\ldots \nu_L}=\frac{1}{L^2}
\bigg ( \sum\limits_{i,j=1}^{L}g^\perp_{\mu_i\nu_j}
O^{\mu_1\ldots\mu_{i-1}\mu_{i+1}\ldots\mu_L}_{\nu_1\ldots
\nu_{j-1}\nu_{j+1}\ldots\nu_L}-
\nonumber \\
 &&  \frac{4}{(2L-1)(2L-3)} \\    &&
\sum\limits_{i<j\atop k<m}^{L}
g^\perp_{\mu_i\mu_j}g^\perp_{\nu_k\nu_m}
O^{\mu_1\ldots\mu_{i-1}\mu_{i+1}\ldots\mu_{j-1}\mu_{j+1}\ldots\mu_L}_
{\nu_1\ldots\nu_{k-1}\nu_{k+1}\ldots\nu_{m-1}\nu_{m+1}\ldots\nu_L}
\bigg )   \nonumber
\end{eqnarray}
The low order projection operators are:
\begin{eqnarray}
&&O=1\qquad\qquad O^\mu_\nu=g^\perp_{\mu\nu}
\nonumber \\
&&O^{\mu\nu}_{\alpha\beta}=\frac 12 \Big (
g^\perp_{\mu\alpha}g^\perp_{\nu\beta}+
g^\perp_{\mu\beta}g^\perp_{\nu\alpha}-\frac 23
g^\perp_{\mu\nu}g^\perp_{\alpha\beta} \Big )
\end{eqnarray}
The scattering of the  two spinless particles  in the partial wave with the total spin $J=L$ (for example a $\pi\pi\to\pi\pi$ transition) is described
as a convolution of the operators
$X^{(L)}(k)$ and $X^{(L)}(q)$ where $k$ and $q$ are relative
momenta before and after the interaction.
\begin{eqnarray}
&X^{(L)}_{\mu_1\ldots\mu_L}(k^\perp)&O^{\mu_1\ldots\mu_L}_
{\nu_1\ldots\nu_L}X^{(L)}_{\mu_1\ldots\mu_L}
(q^\perp) =  \nonumber \\ &&\alpha(L)
\left(\sqrt{k^2_\perp}\sqrt{ q^2_\perp}\right)^{L} P_L(z)\; .
\label{kq_ampl}
\end{eqnarray}
Here $P_L(z)$ are Legendre polynomials and  $z=(k^\perp q^\perp)/(\sqrt{k_\perp^2}\sqrt{q_\perp^2})$ which are, in c.m.s., functions of the cosine of the angle
between initial and final particles.\\
\subsection{The decay of bound system into two pseudoscalar meson states}
The system of the two pseudoscalar particles can form a partial wave with spin and parity $J^P$ where $J=L$ and $P=(-1)^L$. The decay vertex of such system is described by the orbital momentum tensor only:
\begin{eqnarray}
   V^J_{\mu_1\ldots\mu_L}=X^{(L)}_{\mu_1\ldots\mu_L}(k^\perp) ,
\end{eqnarray}
In the two-meson decay the $G$-parity corresponds to the product of the G-parities of the  final particles and the isospin can have values from $|I_1-I_2|$ to $I_1+I_2$ . If the final particles are neutral one  the $C$-parity of the two meson system is equal to the product of the $C$-parities of the final particles. In the case where the final particles are the particle and its own antiparticle $C=(-1)^{L}$.  The identical particles cannot form the bound states with the odd partial waves: such amplitude will be anti-symmetrical and disappear when particles are permuted.  Then for the scattering of two pseudoscalar particles, the combined system can have the quantum numbers listed in Table \ref{2pi}. 
\begin{table}
\caption{\label{2pi} The partial waves in the channel of two pseudoscalar particles.}
\begin{tabular}{lllllll}
\hline
L            & 0& 1&2&3&4 & \\
$\pi^0\pi^0$&  $0^{++}$ & & $2^{++}$& & $4^{++}$ &$\ldots $ \\
$\pi^+\pi^-$&  $0^{++}$&$1^{--}$& $2^{++}$&$3^{--}$&$4^{++}$&$\ldots$  \\
$\pi^\pm\pi^0$& $0^{+}$&$1^{-}$,& $2^{+}$&$3^{-}$&$4^{+}$& $\ldots$ \\
$\eta\eta$&  $0^{++}$&  & $2^{++}$ &  & $4^{++}$ &  $\ldots$ \\ 
$\pi^0\eta$&  $0^{++}$ & $1^{-+}$ &$2^{++}$  & $3^{-+}$ & $4^{++}$ & $\ldots$  \\
$\pi^\pm\eta$ &  $0^{+}$ & $1^{-}$ &$2^{+}$ & $3^{-}$ & $4^{+}$ & $\ldots$ \\
$\eta\eta'$&  $0^{++}$  &$1^{-+}$ &$2^{++}$ & $3^{-+}$ & $4^{++}$ &  $\ldots$ \\
\hline
\end{tabular}
\end{table}
For the two-pion system, the isospin and G parity can be $I^G=0^+,1^+,2^+$, for the $\pi\eta$ system $I^G=1^-$ and for the $\eta\eta\,(\eta\eta')$ system $I^G=0^+$.  In the quark model, where mesons are considered to be bound states of the quark and antiquark, the isospin can be $I=0,1$   $G$-parity is connected with $C$-parity as $G=C(-1)^I$ and states with $CP=-1$ and   with $P=(-1)^J$  are forbidden.

\subsection{The decay of the resonance into three spinless particles. }

Let us consider the composite system decaying into the final three pseudo-scalar mesons via an intermediate two-body system with spin $I_{12}^{G_{12}}J_{12}^{P_{12}C_{12}}$. The tensor which describes the intrinsic spin  is constructed by the convolution of the of the orbital momentum tensor which describes the two-particle intermediate state and the projection operator of three particle system:
\begin{eqnarray}
   S^{(3)}_{\mu_1\ldots\mu_n}(J_{12})= X^{(L_{12})}_{\nu_1\ldots \nu_n}(k_{12}^\perp)O^{\nu_1\ldots\nu_n}_{\mu_1\ldots\mu_n}(P_3).
\end{eqnarray}
Here  $k_1$ and $k_2$ are the momenta of the particles from decay of the intermediate system and 
\begin{eqnarray}
   k_{12\mu}^\perp\! =\!\frac{(k_1+k_2)_\nu}{2} g_{\mu\nu}^{\perp P_{12}}\quad g_{\mu\nu}^{\perp P_{12}}\!=\!g_{\mu\nu}\!-\!\frac{P_{12\mu}P_{12\nu}}{P_{12}^2}
\end{eqnarray}
where $P_{12}=k_1+k_2$, $P_3=k_1+k_2+k_3$ and $n=J_{12}=L_{12}$.  If the total spin of the three-particle system is equal to $J_3$ the orbital momentum between the intermediate state and the spectator particle can be $L_3=|J_3\!-\!J_{12}|,\ldots,J_3\!+\!J_{12}$. If the combination $(L_3\!+\!J_{12}\!-\!J_3)$  is an even number then the decay vertex can be constructed as: 
\begin{eqnarray}
    &&V^{(+1)}_{\mu_1\ldots\mu_{J_3}}(Q_3)=\nonumber \\&&S^{(3)}_{\mu_1\ldots \mu_i\nu_1\ldots\nu_m}(J_{12})X^{(L_3)}_{\nu_1\ldots \nu_m\mu_{i+1}\ldots\mu_{J_3}}(k_3^\perp)~~
    \label{k3_plus}
\end{eqnarray}
where  $k^\perp_3$ is the momentum of the spectator particle orthogonal to the total momentum of the three particle system. The number of convoluted indices $m=(J_{12}+L_3-J_3)/2$ and $i=J_{12}-m$.  The multiindex $Q_3$ matches the principal quantum numbers $Q_3\equiv J_3,L_3,J_{12}$.  If the combination $(J_{12}\!+\!L_3\!-\!J_3)$  is the odd number then the amplitudes is formed by means of the antisymmetric tensor:
\begin{eqnarray}
    &&V^{(-1)}_{\mu_1\ldots\mu_{J_3}}(Q_3)=\varepsilon_{\mu_1\alpha\beta P_3}\times\nonumber \\
    &&S^{(3)}_{\alpha\mu_2\ldots \mu_i\nu_1\ldots\nu_m}(J_{12})\,X^{(L_3)}_{\beta\nu_1\ldots \nu_m\mu_{i+1}\ldots\mu_{J_3}}(k_3^\perp),~~~~~~
    \label{k3_minus}
\end{eqnarray}
$m=(J_{12}+L_3-J_3-1)/2$,  $i=J_{12}-m$ and 
\begin{eqnarray}
    \varepsilon_{\mu\alpha\beta P_3}\equiv\varepsilon_{\mu\alpha\beta\nu} P_{3\,\nu}.
\end{eqnarray}
The final tensor should be symmetrical, traceless and orthogonal to the total momentum of the three particle system which can be done by convolution with the projection operator:
\begin{eqnarray}
   A^{(3,\alpha)}_{\mu_1\ldots\mu_n}(Q_3)= V^{(\alpha)}_{\nu_1\ldots\nu_n}(Q_3)\,O^{\nu_1\ldots\nu_n}_{\mu_1\ldots\mu_n}(P_3).
   \label{a3}
\end{eqnarray}  
Therefore there are two classes of the vertices with $\alpha=+1,-1$ which we will refer below as natural (tensor) and unnatural (pseudo-tensor) structures. The parity of these states are defined as:
\begin{eqnarray}
\quad P=(-1)^{L_3+J_{12}}\prod\limits_{i=1}^3 P_i,
\end{eqnarray}
where $P_i$ denote the parities of the final states. Table \ref{3pi} enumerates the potential states for the final state $3\pi^0$, considering the intermediate states of the scalar and tensor.
\begin{table}[pt]
\caption{\label{3pi} List of the partial wave states produced in $3\pi^0$ channel. The partial waves are denoted as $J^{PC}_{m \alpha}$  (see eqs.(\ref{k3_plus},\ref{k3_minus},\ref{a3})) . The standard $q\bar q$ states in $3\pi^0$ channel have iospin $I=1$.}
\renewcommand{\arraystretch}{1.3}
\begin{tabular}{l|c|c}
\hline
 & $J_{12}=0$ & $J_{12}=2$ \\
 \hline
$L_3 = 0$ & $0^{-+}_{0+}$ & $2^{-+}_{0+}$ \\
$L_3 = 1$ & $1^{++}_{0+}$ & $1^{++}_{1+}\ 2^{++}_{0-}\ 3^{++}_{0+}$ \\
$L_3 = 2$ & $2^{-+}_{0+}$ & $0^{-+}_{2+}\ 1^{-+}_{1-}\ 2^{-+}_{1+}\ 3^{-+}_{0-}\ 4^{-+}_{0+}$\\
$L_3 = 3$ & $3^{++}_{0+}$& $1^{++}_{2+}\ 2^{++}_{1-}\ 3^{++}_{1+}\ 4^{++}_{0-}\ 5^{++}_{0+}$\\
\hline
\end{tabular}
\renewcommand{\arraystretch}{1.0}
\end{table}

\subsection{The decay of the resonance into four spinless particles through three body intermediate state. }

This case is very similar to the one described above. If the 3-body bound state  has quantum numbers $I_{3}^{G_{3}}J_{3}^{P_{3}C_{3}}$ then the intrinsic spin can be described as projection of the three body amplitude eq.\ref{a3} into 4-particle system:
\begin{eqnarray}
   S^{(4,\alpha)}_{\mu_1\ldots\mu_{J_3}}(Q_3)= A^{(3,\alpha)}_{\nu_1\ldots \nu_{J_3}}(Q_3)O^{\nu_1\ldots\nu_{J_3}}_{\mu_1\ldots\mu_{J_3}}(P),
\end{eqnarray} 
where $P=k_1+k_2+k_3+k_4$.  The orbital momentum between the 3-body state and a spectator particle can be $L_4=|J_4\!-\!J_3|,\ldots,J_4\!+\!J_{3}$. If the combination $(L_4\!+\!J_3\!-\!J_4)$  is an even number, then the decay vertex can be written as convolution of the intrinsic spin tensor with orbital momentum tensor : 
\begin{eqnarray}
    V^{(+1,\alpha)}_{\mu_1\ldots\mu_{J_4}}(Q_4)=S^{(4,\alpha)}_{\mu_1\ldots \mu_i\nu_1\ldots\nu_m}(Q_{3})X^{(L_4)}_{\nu_1\ldots \nu_m\mu_{i+1}\ldots\mu_{J_4}}(k_4^\perp)\nonumber \\
    \end{eqnarray}
where $k^\perp_4$ is the momentum of the spectator particle orthogonal to the momentum of the four particle system. The number of convoluted indices is equal to $m=(J_3+L_4-J_4)/2$  and $i=J_3\!-\!m$. The multiindex $Q_4$ lists all relevant quantum numbers $Q_4=J_4,L_4,J_3,L_3,J_{12}$.  If the combination $(J_{3}\!+\!L_4\!-\!J_4)$  is the odd number:
\begin{eqnarray}
   && V^{(-1,\alpha)}_{\mu_1\ldots\mu_{J_4}}(Q_4)=\varepsilon_{\mu_1\eta\beta P}\times\nonumber \\
   && S^{(4,\alpha)}_{\eta\mu_2\ldots \mu_i\nu_1\ldots\nu_m}(Q_3)X^{(L_4)}_{\beta\nu_1\ldots \nu_m\mu_{i+1}\ldots\mu_{J_4}}(k_4^\perp),~~~~~~
\end{eqnarray}
$m=(J_{3}+L_4-J_4-1)/2$ and $i=J_3\!-\!m$. The final tensor should be symmetrical, traceless and orthogonal to the total momentum of the four particle system which can be done by means the  projection operator:
\begin{eqnarray}
   A^{(4,\beta,\alpha)}_{\mu_1\ldots\mu_{J_4}}(Q_4)= V^{(\beta, \alpha)}_{\nu_1\ldots\nu_{J_4}}(Q_4)O^{\nu_1\ldots\nu_{J_4}}_{\mu_1\ldots\mu_{J_4}}(P),
   \end{eqnarray}  
where $\beta,\alpha\!=\!-1,+1$.  Thus we have 4 classes of amplitudes:
\begin{eqnarray}
  &+1,+1:J_3\!+\!L_4\!-\!J_4 \!=\!2n~~~~ &J_{12}\!+\!L_3\!-\!J_3\!=\!2m\\
  &+1,-1: J_3\!+\!L_4\!-\!J_4 \!=\!2n~~~~ &J_{12}\!+\!L_3\!-\!J_3\!=\!2m\!+\!1~~~~~~~\\
  &-1,+1:J_3\!+\!L_4\!-\!J_4 \!=\!2n\!+\!1 &J_{12}\!+\!L_3\!-\!J_3\!=\!2m\\
  &-1,-1:J_3\!+\!L_4\!-\!J_4 \!=\!2n\!+\!1 &J_{12}\!+\!L_3\!-\!J_3\!=\!2m\!+\!1~~~~~~~
\end{eqnarray}

The parity of the 4-particle state are defined as: 
\begin{eqnarray}
\quad P=(-1)^{L_4+L_3+J_{12}}\prod\limits_{i=1}^4 P_i. 
\end{eqnarray}  
The partial wave amplitudes in the $4\pi^0$ channel in the case of cascade decay are listed in Table~\ref{t4_cascade} for the intermediate three body states up to $J_3\!=\!2$. The exotic states which can be produced in the $4\pi^0$ and $3\pi^0$ channels are also included in this Table.



\begin{table}[pt]
\caption{\label{t4_cascade}List of the partial wave amplitudes $J^{PC}_{n \beta}$ state for cascade decay into $4\pi^0$. }
\renewcommand{\arraystretch}{1.3}
\begin{tabular}{l|c|c|c}
\hline
 $J^{PC}_3$& $0^{-+}$ & $1^{++}$ & $1^{-+}$ \\
 \hline
$L_4 = 0$ & $0^{++}_{0+}$& $1^{-+}_{0+}$ & $1^{++}_{0+}$ \\
$L_4 = 1$ & $1^{-+}_{0+}$& $0^{++}_{1+}\ 1^{++}_{0-}\ 2^{++}_{0+}$& $0^{++}_{0+}\ 1^{++}_{0-}\ 2^{++}_{0+}$
\\
$L_4 = 2$ & $2^{++}_{0+}$& $1^{-+}_{1+}\ 2^{-+}_{0-}\ 3^{-+}_{0+}$&$1^{-+}_{1+}\ 2^{-+}_{0-}\ 3^{-+}_{0+}$\\
\hline
 $J^{PC}_3$& \multicolumn{2}{c|}{$2^{++}$} &  $2^{-+}$\\
 \hline
$L_4 = 0$ & \multicolumn{2}{c|}{$2^{-+}_{0+}$} & $2^{++}_{0+}$ \\
$L_4 = 1$ & \multicolumn{2}{c|}{$1^{++}_{0+}\ 2^{++}_{0-}\ 3^{++}_{0+}$}
&$1^{-+}_{1+}\ 2^{-+}_{0-}\ 3^{-+}_{0+}$\\
$L_4 = 2$ & \multicolumn{2}{c|}{$0^{-+}_{2+}\ 1^{-+}_{1-}\ 2^{-+}_{1+}\ 3^{-+}_{0-}\ 4^{-+}_{0+}$}&  $0^{++}_{2+}\ 1^{++}_{1-}\ 2^{++}_{1+}\ 3^{++}_{0-}\ 4^{++}_{0+}$\\
\hline
\end{tabular}

\renewcommand{\arraystretch}{1.0}
\end{table}


\subsection{The decay of the resonance into two resonances decaying into two spinless particles}

Consider the decay of the resonance into two resonances with spin $J_{12}$ and $J_{34}$ that decay into two spinless particles with momenta $k_1,k_2$ and $k_3,k_4$, respectively. The decay of these two states is described with $X^{(J_{12})}_{\mu_1\ldots\mu_{J_{12}}}(k^\perp_{12})$ and $X^{(J_{34})}_{\mu_1\ldots\mu_{J_{34}}}(k^\perp_{34})$ tensors, where
\begin{eqnarray}
   k^\perp_{ij\mu}\!=\!\frac{k_{i\nu}\!-\!k_{j\nu}}{2} g^{\perp P_{ij}}_{\mu\nu} 
   \quad g^{\perp P_{ij}}_{\mu\nu}=g_{\mu\nu} \!-\!\frac{P_{ij\mu} P_{ij\nu}}{P_{ij}^2},
\end{eqnarray} 
where $P_{ij}=k_i+k_j$. The intrinsic spin can have values $S=|J_{12}\!-\!J_{34}|,\ldots, J_{12}\!+\!J_{34}$.  If $J_{12}+J_{34}-S$ = 2m, the tensor that describes this spin state can be formed by convolution of the $m$ indices.
\begin{eqnarray}
&&V^{(+1)}_{\mu_1\ldots\mu_S}(Q_{S})=
\nonumber\\ 
&&X^{(J_{12})}_{\mu_1\ldots\mu_{J_{12}-m}\nu_1\ldots\nu_m}(k_{12}^\perp) X^{(J_{34})}_{\nu_1\ldots\nu_m\mu_{J_{12}-m+1}\ldots\mu_{S}}(k_{34}^\perp) ~~~~~
\end{eqnarray}
where $Q_{S}=S,J_{12},J_{34}$. If the difference between intrinsic spin and the sum of spin resonances is an odd number $J_{12}\!+\!J_{34}\!-\!S\!=\!2m\!+\!1$ the vertex is formed by means of the antisymmetric tensor. 
\begin{eqnarray}
&&V^{(-1)}_{\mu_1\ldots\mu_S}(Q_{S})=\varepsilon_{\mu_1\eta\beta P }\, X^{(J_{12})}_{\eta\mu_2\ldots\mu_{J_{12}-m}\nu_1\ldots\nu_m}(k_{12}^\perp)
\times
\nonumber\\ 
&& 
X^{(J_{34})}_{\beta\nu_1\ldots\nu_m\mu_{J_{12}-m+1}\ldots\mu_{S}}(k_{34}^\perp),~~~ 
\end{eqnarray}
where $P=k_1\!+\!k_2\!+\!k_3\!+\!k_4$.
The symmetrization of indices and traceless property can be satisfied in the standard way by convolution with the projection operator:
\begin{eqnarray}
  S^{(22,\alpha)}_{\mu_1\ldots\mu_S}(Q_{S})=V^{(\alpha)}_{\nu_1\ldots\nu_S}(Q_{S})O ^{\nu_1\ldots\nu_S}_{\mu_1\ldots\mu_S}(P), 
\end{eqnarray}
If the spin of the 4-particle partial wave is equal $J_4$, the orbital momentum between the two resonances can be $L_4=|J_4\!-\!S|,\ldots,J_4\!+\!S$. If the combination $(L_4\!+\!S\!-\!J_4)=2n$  is the even number then the decay vertex can be written as: 
\begin{eqnarray}
    &&V^{(+1,\alpha)}_{\mu_1\ldots\mu_{J_4}}(Q_{22})=\nonumber \\&&S^{(22,\alpha)}_{\mu_1\ldots \mu_{S-n}\nu_1\ldots\nu_n}(Q_{S})X^{(L_4)}_{\nu_1\ldots \nu_n\mu_{S-n+1}\ldots\mu_{J_4}}(k^\perp)~~
\end{eqnarray}
where  $k^\perp$ is the relative momentum between two intermediate resonances:
\begin{eqnarray}
   k_\mu^\perp =\frac 12(k_1+k_2-k_3-k_4)_\nu g^\perp_{\mu\nu}
   \label{k22}
\end{eqnarray}
and multi-index $Q_{22}=J_4,L_4,S,J_{12},J_{34}$.  If the combination $(J_{3}\!+\!L_4\!-\!J_4)$  is the odd number:
\begin{eqnarray}
   && V^{(-1,\alpha)}_{\mu_1\ldots\mu_{J_4}}(Q_{22})=\varepsilon_{\mu_1\eta\beta P}\times\nonumber \\
   && S^{(22,\alpha)}_{\eta\mu_2\ldots \mu_{S-n}\nu_1\ldots\nu_n}(Q_S)X^{(L_4)}_{\beta\nu_1\ldots \nu_n\mu_{S-n+1}\ldots\mu_{J_4}}(k^\perp),~~~~~~
\end{eqnarray}
$n\!=\!(S\!+\!L_4\!-\!J_4\!-\!1)/2$. The final tensor should be symmetrical, traceless and orthogonal to the total momentum of the four particle system which can be done with projection operator:
\begin{eqnarray}
   A^{(22,\beta,\alpha)}_{\mu_1\ldots\mu_{J_4}}(Q_{22})= V^{(\beta, \alpha)}_{\nu_1\ldots\nu_{J_4}}(Q_{22})O^{\nu_1\ldots\nu_{J_4}}_{\mu_1\ldots\mu_{J_4}}(P),
   \end{eqnarray}    
   Thus we have 4 classes of amplitudes:
\begin{eqnarray}
  &+1,+1:S\!+\!L_4\!-\!J_4 \!=\!2n~~~~ &J_{12}\!+\!J_{34}\!-\!S\!=\!2m\\
  &+1,-1: S\!+\!L_4\!-\!J_4 \!=\!2n~~~~ &J_{12}\!+\!J_{34}\!-\!S\!=\!2m\!+\!1~~~~~~~\\
  &-1,+1:S\!+\!L_4\!-\!J_4 \!=\!2n\!+\!1 &J_{12}\!+\!J_{34}\!-\!S\!=\!2m\\
  &-1,-1:S\!+\!L_4\!-\!J_4 \!=\!2n\!+\!1 &J_{12}\!+\!J_{34}\!-\!S\!=\!2m\!+\!1~~~~~~~
\end{eqnarray}
The parity of the 4-particle state are defined as: 
\begin{eqnarray}
\quad P=(-1)^{L_4+J_{12}+J_{34}}\prod\limits_{i=1}^4 P_i. 
\end{eqnarray}

\begin{table}[pt]
\caption{ List of possible spin combinations $S^{P}_{m \alpha}$ states ($C\!=\!+1$) in four particle decay via two resonance decays. }
\renewcommand{\arraystretch}{1.3}
\begin{tabular}{l|c|c|c|c}
\hline
  $J^{PC}_{12}$ &$0^{++}$& $0^{++}$ & $2^{++}$ & $2^{++}$\\
  $J^{PC}_{34}$ &$0^{++}$& $2^{++}$ & $2^{++}$ & $4^{++}$\\
 \hline
$S^{P}_{m\alpha}$& $0^{+}_{0+}$&   $2^{+}_{0+}$& $0^{+}_{2+}\ 1^{+}_{1-}\ 2^{+}_{1+}\ 3^{+}_{0-}\ 4^{+}_{0+}$& $2^{+}_{2+}\ 3^{+}_{1-}\ 4^{+}_{1+}\ 5^{+}_{0-}\ 6^{+}_{0+}$ \\
\hline
\end{tabular}
\renewcommand{\arraystretch}{1.0}
\end{table}

\begin{table}[pt]
\caption{\label{22pi}List of states $J^{PC}_{n\beta}$ in $4\pi^0$ channel decaying into two resonances forming intrinsic spin $S^{PC}$.}
\renewcommand{\arraystretch}{1.3}
\begin{tabular}{l|c|c|c}
\hline
 $S^{PC}$     & $0^{++}$ & $1^{++}$ & $2^{++}$   \\ 
\hline 
$L_4=0$ & $0^{++}_{0+}$ & $1^{++}_{0+}$ & $2^{++}_{0+}$   \\
$L_4=1$ & $1^{-+}_{0+}$ & $0^{-+}_{1+}\ 1^{-+}_{0-}\ 2^{-+}_{0+}$ & $1^{-+}_{1+}\ 2^{-+}_{0-}\ 3^{-+}_{0+}$  \\
$L_4=2$ & $2^{++}_{0+}$ & $1^{++}_{1+}\ 2^{++}_{0-}\ 3^{++}_{0+}$ & $0^{++}_{2+}\ 1^{++}_{1-}\ 2^{++}_{1+}\ 3^{++}_{0-}\ 4^{++}_{0+}$  \\
\hline
 $S^{PC}$     &  \multicolumn{2}{c|}{$3^{++}$} & $4^{++}$   \\ 
\hline 
$L_4=0$ & \multicolumn{2}{c|}{$3^{++}_{0+}$} & $4^{++}_{0+}$  \\
$L_4=1$ & \multicolumn{2}{c|}{$2^{-+}_{1+}\ 3^{-+}_{0-}\ 4^{-+}_{0+}$} & $3^{-+}_{1+}\ 4^{-+}_{0-}\ 5^{-+}_{0+}$  \\
$L_4=2$ & \multicolumn{2}{c|}{$1^{++}_{2+}\ 2^{++}_{1-}\ 3^{++}_{1+}\ 4^{++}_{0-}\ 5^{++}_{0+}$} & $2^{++}_{2+}\ 3^{++}_{1-}\ 4^{++}_{1+}\ 5^{++}_{0-}\ 6^{++}_{0+}$  \\
\hline
\end{tabular}
\renewcommand{\arraystretch}{1.0}
\end{table}

\subsection{The production of the states decaying into 4 pseudoscalar mesons.}

A partial wave amplitude is a scalar value.  The  convolution of the tensors with antisymmetric tensor creates a pseudo-tensor  structure (unnatural). Thus the scalar value should be either convolution of any number of tensors (natural structures) or a convolution of the even number of unnatural structures. Therefore if the produced state is a natural state the decay amplitude can have  only  $(++)$ or $(--)$ structures.  If the produced state is an unnatural state, the decay should be described either by the $(+-)$  or  $(-+)$ combination.    


\subsubsection{The production of the state from pomeron-pomeron, $f_0 f_0$ or $\pi^0\pi^0$   collision}

Consider the production of resonances in central collision reactions.  In that case, the states can be produced, for example,  from the pomeron-pomeron  collision. The similar mechanism is responsible for the production of resonances in two-scalar or two neutral pseudo-scalar mesons.  In such processes, only states with $G\!=\!+1$ and $C\!=\!+1$ are produced. In the standard quark model only resonances with isospin 0 and even spin can be produced in such reaction:
$I^G J^{PC}=0^+J^{++}$ and $J=2n$. The production vertex is described as the orbital momentum tensor $X^{(J_4)}_{\mu_1\ldots\mu_{J_4}}(q^\perp)$  where $q$ is the relative momenta of the collided particles. Then the partial wave amplitude for the partial wave with spin $J_4$  has the form:
\begin{eqnarray}
A^{(t\alpha\beta)}_{J_4}=X ^{(J_4)}_{\mu_1\ldots\mu_{J_4}}(q^\perp)A ^{(t\alpha\beta)}_ {\mu_1\ldots\mu_{J_4}}(Q_{t}),
\end{eqnarray}
where index $t$ describes topologies $t=4$ (cascade) and $t=22$ (decay into two resonances). 
Consider the case with the final state $4\pi^0$ (or $4\eta$). Then only resonances with $I^GJ^{PC}=0^+J_{ij}^{++}$ where $J_{ij}=2n$ can be produced in the two-particle channel. 
If we have two scalar resonances $J_{12}=J_{34}=0$ in the final states only $++$ amplitudes are possible:
\begin{eqnarray}
  &A^{(22++)}_{J_4}&=X ^{(J_4)}_{\mu_1\ldots\mu_{J_4}}(q^\perp) O^ {\mu_1\ldots\mu_{J_4}}_{\nu_1\ldots\nu_{J_4}}(P) X ^{(J_4)}_{\nu_1\ldots\nu_{J_4}}(k^\perp)\nonumber \\&&=X ^{(J_4)}_{\mu_1\ldots\mu_{J_4}}(q^\perp) X ^{(J_4)}_{\mu_1\ldots\mu_{J_4}}(k^\perp) 
\end{eqnarray} we obtain the standard Legendre polynomial dependence  in the rest system of the 4-particle  state (see eq.\ref{kq_ampl}).  In the cascade topology the amplitudes have $++$ or $--$ signatures. The examples of other amplitudes are given in the Appendix.

\subsection{The production of the states decaying into 4 pseudoscalar particles in the $J/\Psi$ radiative decays.}

The states produced in radiative decay $J/\Psi$ can be natural or unnatural states.  All states have isospin 0, $G\!=\!+1$ and $C\!=\!+1$. This process can be considered as the production of resonances in collisions $J/\Psi\gamma$. In this case  the intrinsic spin can have values $S=0,1,2$, and the corresponding spin tensors are:
\begin{eqnarray}
 &S^{0}= g_{\mu\nu}\epsilon^\Psi_\mu \epsilon^{\gamma *}_\nu &\qquad S=0\nonumber \\
 &S^1_\eta=\varepsilon_{\eta\mu\nu P}\epsilon^\Psi_\mu \epsilon^{\gamma *}_\nu &\qquad S=1 \nonumber \\
 &S^2_{\eta\xi}=\epsilon^\Psi_\mu \epsilon^{\gamma *}_\nu O^{\mu\nu}_{\eta\xi}(P)& \qquad S=2
\end{eqnarray}
In the spin-orbital basis ($SL$) the partial waves can be described as $^{2S+1}L_J$.  In this basis, the partial waves with $S\!=\!0$  have $L\!=\!J$ and parity $P=(-1)^J$. These production vertices  are described as:
\begin{eqnarray}
    V^0_{\mu_1\ldots\mu_J}=S^0X^{(J)}_{\mu_1\ldots\mu_J}(q^\perp)\qquad^1J_J
\end{eqnarray}
and all amplitudes are the natural ones.  The natural amplitudes can be constructed with intrinsic spin 1 only with $L=J$ ($^3L_J$):
\begin{eqnarray}
  V^{1,0}_{\mu_1\ldots\mu_J}=\varepsilon_{\mu_1\eta\xi P}S^1_\eta X^{(J)}_{\xi\mu_2\ldots\mu_J}(q^\perp)\qquad ^3J_J. 
\end{eqnarray}
This vertex can be symmetrized and traceless by convolution with the projection operator. However, we already applied this operator to the decay vertices, and it is not necessary to apply it to the production vertices. For the amplitudes with $S\!=\!2$ the natural amplitudes are produced with $L\!=\!J\!-\!2,J,J\!+\!2$ :
\begin{eqnarray}
&V^{5,-2}_{\mu_1\ldots\mu_J}=&S^2_{\mu_1\mu_2} X^{(J-2)}_{\mu_3\ldots\mu_J}(q^\perp)\qquad^5(J\!-\!2)_J\nonumber \\
&V^{5,0}_{\mu_1\ldots\mu_J}=&S^2_{\mu_1\eta} X^{(J)}_{\eta\mu_2\ldots\mu_J}(q^\perp)\qquad^5J_J\nonumber \\
&V^{5,+2}_{\mu_1\ldots\mu_J}=&S^2_{\xi\eta} X^{(J+2)}_{\xi\eta\mu_1\ldots\mu_J}(q^\perp)\qquad^5(J\!+\!2)_J
\end{eqnarray}
Recall that the decay of the natural states can be described with $--$ or $++$ amplitudes in the decay into 4 pseudoscalar mesons. In the gauge-invariant limit for the partial waves with $J\ge 2$ only three amplitudes are linearly independent. For $J=0$ only one amplitude is linearly independent and for $J=1$ we have two linearly independent amplitudes (see Appendix).

The unnatural amplitudes are formed with intrinsic spin $S=1,2$ and the orbital momentum $L=J\!-\!1$ and $L=J\!+\!1$: 
\begin{eqnarray}
  V^{3,-1}_{\mu_1\ldots\mu_J}=&S^1_{\mu_1} X^{(J-1)}_{\mu_2\ldots\mu_J}(q^\perp)&\quad ^3(J\!-\!1)_J\nonumber \\  
  V^{3,+1}_{\mu_1\ldots\mu_J}=&S^1_{\eta} X^{(J+1)}_{\eta\mu_1\ldots\mu_J}(q^\perp)&\quad ^3(J\!+\!1)_J\nonumber \\ 
  V^{5,-1}_{\mu_1\ldots\mu_J}=&\varepsilon_{\mu_1\eta\xi P}S^2_{\eta\mu_2} X^{(J-1)}_{\xi\mu_3\ldots\mu_J}(q^\perp)&\quad ^5(J\!-\!1)_J\nonumber \\  
  V^{5,+1}_{\mu_1\ldots\mu_J}=&\varepsilon_{\mu_1\eta\xi P}S^2_{\eta\chi} X^{(J+1)}_{\xi\chi\mu_2\ldots\mu_J}(q^\perp)&\quad ^5(J\!+\!1)_J ~~~
\end{eqnarray}

In the gauge-invariant limit for the partial waves with $J\ge 2$ only three amplitudes are linearly independent. For $J=0$ only one amplitude is linearly independent and for $J=1$ we have two linearly independent amplitudes. The details are given in the Appendix together with examples of the amplitudes for the radiative $J/\psi$ decay into 4$\pi^0$. 

\section{Conclusion }
 We have developed the formalism for the partial wave analysis of the data with four pseudoscalar meson states. The method is covariant and can be directly applied to the event-by-event analysis of the data, for example, in the framework of the maximum likelihood method. In particular, we consider the $4\pi^0$ production in the radiative decay $J/\Psi$ and the production of the four pseudo-scalar mesons in the central collision.

\section{Acknowledgments}
The paper is supported by the RNF grant 24-22-00322.
\begin{table}[pt]
\caption{\label{psi}List of partial waves produced in radiative $J/\Psi$ decay. The partial waves which became linear dependent from other partial waves are shown in third column. The states which are forbidden in the quark model are not listed.}
\renewcommand{\arraystretch}{1.3}
\begin{tabular}{l|c|c}
\hline
$0^{++}$ & $^{1}S_{0}$ &  $^{5}D_{0}$ \\
$2^{++}$ & $^{1}D_{2}$ $^{3}D_{2}$ $^{5}S_{2}$ & $^{5}D_{2}$ $^{5}G_{2}$  \\
$4^{++}$ & $^{1}G_{4}$ $^{3}G_{4}$ $^{5}D_{4}$ & $^{5}G_{4}$ $^{5}I_{4}$\\
\hline 
$0^{-+}$ & $^{3}P_{0}$ &\\
$1^{++}$ & $^{3}S_{1}$ $^{3}D_{1}$&  $^{5}D_{1}$\\
$2^{-+}$ & $^{3}P_{2}$ $^{5}P_{2}$ $^{3}F_{2}$ & $^{5}F_{2}$ \\
$3^{++}$ & $^{3}D_{3}$ $^{3}G_{3}$ $^{5}D_{3}$ &  $^{5}G_{3}$ \\
\hline
\end{tabular}
\renewcommand{\arraystretch}{1.0}
\label{jpsilist}
\end{table} 

\section{Appendix}

\subsection{Linearly dependent vertices in the gauge limit for the $J/\Psi$ radiative decay}

Consider the general structure of $J/\Psi$ radiative decay. The production of the state with spin $J$ can be described as a convolution projection operator with vertex:
\begin{eqnarray}
 O^{\nu_1\ldots\nu_J}_{\mu_1\ldots\mu_J}(P)V_{\mu_1\ldots\mu_J} (q^\perp)  
\end{eqnarray}
where $q^\perp$ is momentum of $J/\Psi$ orthogonal to the momentum of the resonance. Taking into account the orthogonality and traceless properties of the projection operator the general structure of the vertex for the natural states is:
\begin{eqnarray}
  &&V_{\mu_1\ldots\mu_J} =(\epsilon ^{\Psi}\epsilon ^{\gamma*})q^\perp_{\mu_1}\ldots q^\perp_{\mu_J}F_1+\epsilon ^{\Psi}_{\mu_1}\epsilon ^{\gamma*}_{\mu_2}q^\perp_{\mu_3}\ldots q^\perp_{\mu_J}F_2\nonumber \\
  &&+\epsilon ^{\Psi}_{\mu_1}(\epsilon ^{\gamma*}q^\perp) q^\perp_{\mu_2}\ldots q^\perp_{\mu_J}F_3+(\epsilon ^{\Psi}q^\perp)\epsilon ^{\gamma*}_{\mu_1} q^\perp_{\mu_2}\ldots q^\perp_{\mu_J}F_4\nonumber \\&&+(\epsilon ^{\Psi}q^\perp)(\epsilon ^{\gamma*}q^\perp) q^\perp_{\mu_1}\ldots q^\perp_{\mu_J}F_5 
\end{eqnarray} where $F_i$ are scalar functions. 
Remember that polarization vector of the particle is orthogonal to its momentum, and we can use $q^\perp$ as universal momentum in convolution with both polarization vectors.  In the gauge limit the polarization vector of photon is orthogonal to all momenta in the vertex and therefore $(\epsilon ^{\gamma*}q^\perp)=0$. As a consequence, only structures with $F_1,F_2$ and $F_4$ will contribute to the decay. In the case of the scalar state ($J=0$) the vertex has only structures $F_1,F_5$  and only $F_1$ is not zero in the gauge limit. For the vector state ($J=1$) only structures $F_1,F_3,F_4,F_5$   can contribute to the vertex and only $F_1$ and $F_4$ survive in the gauge limit.

For the production of unnatural states the vertex has the structure:
\begin{eqnarray}
 &&V_{\mu_1\ldots\mu_J} =\varepsilon_{q^\perp\alpha\beta P}\epsilon ^{\Psi}_\alpha \epsilon ^{\gamma*}_\beta q^\perp_{\mu_1}\ldots q^\perp_{\mu_J}F_1
 \nonumber \\
 &&+\varepsilon_{\mu_1\alpha\beta P}\epsilon ^{\Psi}_\alpha\epsilon ^{\gamma*}_\beta q^\perp_{\mu_2}\ldots q^\perp_{\mu_J}F_2  \nonumber \\
 &&+\varepsilon_{\mu_1\alpha q^\perp P}(\epsilon ^{\Psi}_\alpha \epsilon ^{\gamma*}_{\mu_2}+\epsilon ^{\Psi}_{\mu_2} \epsilon ^{\gamma*}_{\alpha}) q^\perp_{\mu_3}\ldots q^\perp_{\mu_J}F_3\nonumber \\
&& +\varepsilon_{\mu_1\alpha q^\perp P}\left(\epsilon ^{\Psi}_\alpha (\epsilon ^{\gamma*}q^\perp)+ (\epsilon ^{\Psi}q^\perp) \epsilon ^{\gamma*}_\alpha\right ) q^\perp_{\mu_2}\ldots q^\perp_{\mu_J}F_4~~~~~~
\label{unnatural}
\end{eqnarray}
For $J^{PC}=0^{-+}$ state  only one structure (with $F_1$) exists. In the gauge limit $\epsilon ^{\gamma *}_\alpha$  is orthogonal to momentum $q^\perp$.  Decompose the polarization vector $\epsilon ^{\Psi}_\alpha$ into components parallel to $q^\perp$, parallel to $P$, and orthogonal to both $q^\perp$ and $P$.  Then: 
\begin{eqnarray}
 \epsilon ^{\Psi}_\alpha&&=C q^\perp_\alpha +\epsilon ^{\perp\perp}_\alpha+\delta P_\alpha
\nonumber \\
 \varepsilon_{\mu_1\alpha\beta P}\epsilon ^{\Psi}_\alpha&&=C  \varepsilon_{\mu_1 q_\perp\beta P}+\varepsilon_{\mu_1 \alpha\beta P}\epsilon ^{\perp\perp}_\alpha \\
 \varepsilon_{\mu_1\alpha\beta P}\epsilon ^{\Psi}_\alpha\epsilon ^{\gamma*}_\beta &&=C\varepsilon_{\mu_1 q^\perp\beta P}\epsilon ^{\gamma*}_\beta + \varepsilon_{\mu_1 \alpha\beta P}\epsilon^{\perp\perp}_\alpha \epsilon ^{\gamma*}_\beta \nonumber 
\end{eqnarray}
As both $\epsilon^{\perp\perp}_\alpha, \epsilon ^{\gamma*}_\beta$ are orthogonal to momentum $q^\perp$ and $P$, then
\begin{eqnarray}
 \varepsilon_{\mu_1 \alpha\beta P}\epsilon^{\perp\perp}_\alpha \epsilon ^{\gamma*}_\beta = B q^\perp_{\mu_1}
\end{eqnarray}
Then we obtain the following structures for the vertex:
\begin{eqnarray}
 &&V_{\mu_1\ldots\mu_J} =B (q_\perp^2 F_1+F_2)q^\perp_{\mu_1}\ldots q^\perp_{\mu_J}
 \nonumber \\
 &&-(C\,F_2+(\epsilon ^{\Psi}q^\perp)F_4)\varepsilon_{\mu_1 \beta q^\perp P}\epsilon ^{\gamma*}_\beta q^\perp_{\mu_2}\ldots q^\perp_{\mu_J} \nonumber \\
&&+F_3\varepsilon_{\mu_1\alpha q^\perp P}(\epsilon ^{\Psi}_\alpha \epsilon ^{\gamma*}_{\mu_2}+\epsilon ^{\Psi}_{\mu_2} \epsilon ^{\gamma*}_{\alpha}) q^\perp_{\mu_3}\ldots q^\perp_{\mu_J}
 \end{eqnarray}
It is seen that amplitudes $F_1,F_2,F_4$ have only two independent structures. This means that only two of them are linearly independent. This corresponds to the case $J^P=1^+$ where the structure with $F_3$ cannot be produced. Thus, for $J\ge 2$, only three partial waves are linearly independent for every unnatural state. The list of the lowest partial waves is given in Table \ref{psi}.

\subsection{The decay of the resonances into $4\pi^0$ final state.}
Let us list  the basic momenta and tensors: 

\begin{eqnarray}
&& P_{12\, \mu} = (k_1 + k_2)_{\mu}\, , \quad
g_{\mu\nu}^{\perp P_{12}} = g_{\mu\nu} -\frac{P_{12\, \mu}P_{12\, \nu}}{P_{12}^2}\,, \nonumber \\
&& k_{12\, \mu}^{\perp} = {\textstyle\frac12} (k_1 - k_2)_{\nu} g_{\nu\mu}^{\perp P_{12}}\nonumber \\
&& P_{34\, \mu} = (k_3 + k_4)_{\mu}\, , \quad
g_{\mu\nu}^{\perp P_{34}} = g_{\mu\nu} -\frac{P_{34\, \mu}P_{34\, \nu}}{P_{34}^2}\,, \nonumber \\
&& k_{34\, \mu}^{\perp} = {\textstyle\frac12} (k_3 - k_4)_{\nu} g_{\nu\mu}^{\perp P_{34}}\nonumber
\end{eqnarray}
\begin{eqnarray}
&& P_{3\, \mu} = (P_{12} +k_3)_\mu\, , \quad
g_{\mu\nu}^{\perp P_{3}} = g_{\mu\nu} -\frac{P_{3\, \mu}P_{3\, \nu}}{P_{3}^2}\,, \nonumber \\
&& k_{3\, \mu}^{\perp} = 1/2 (k_3-P_{12} )_{\nu} g_{\nu\mu}^{\perp P_{3}}\nonumber \\
&& P_{\mu} = (P_{3} +k_4)_\mu\, , \quad
g_{\mu\nu}^{\perp} = g_{\mu\nu} -\frac{P_{\mu}P_{\nu}}{P^2}\,, \nonumber \\
&& k_{4\,\mu}^{\perp} = {\textstyle\frac12} (k_4-P_3 )_{\nu} g_{\nu\mu}^{\perp} \nonumber \\
&& k_{\mu}^{\perp} = {\textstyle\frac12} (k_1+k_2- k_3-k_4)_{\nu} g_{\nu\mu}^{\perp} \nonumber \\
&& O^{\nu_1\ldots\nu_J}_{\mu_1\ldots\mu_J} \equiv O^{\nu_1\ldots\nu_J}_{\mu_1\ldots\mu_J}(P)
\end{eqnarray}

Consider the decay of states into the $4\pi^0$ channel.  Let us start from decay via two intermediate resonances for natural states :
\begin{eqnarray}
& J^{PC}&\to J^{PC}_{n,\,\beta}(L_4 (S^{PC}_{m\alpha}\to f_{J_{12}}f_{J_{34}} ))\nonumber \\
&0^{++}&\to 0^{++}_{0,+}(0 (0^{++}_{0+}\to f_{0}f_{0} ))\nonumber \\
&V^{(0+,1)}&=1  \\
&0^{++}&\to 0^{++}_{2,+}(2 (2^{++}_{0+}\to f_{0}f_{2} ))\nonumber \\
&V^{(0+,2)}&=X^{(2)}_{\alpha\beta}(k^\perp) X^{(2)}_{\alpha\beta}(k_{34}^\perp) \\
&0^{++}&\to 0^{++}_{0,+}(0 (0^{++}_{2+}\to f_{2}f_{2} ))\nonumber \\
&V^{(0+,3)}&=X^{(2)}_{\alpha\beta}(k_{12}^\perp) X^{(2)}_{\alpha\beta}(k_{34}^\perp) \\
&0^{++}&\to 0^{++}_{0,+}(2 (2^{++}_{1+}\to f_{2}f_{2} ))\nonumber \\
&V^{(0+,4)}&=X^{(2)}_{\alpha\beta}(k^\perp)X^{(2)}_{\alpha\xi}(k_{12}^\perp) X^{(2)}_{\beta\xi}(k_{34}^\perp) \\
&2^{++}&\to 2^{++}_{0,+}(2 (0^{++}_{0+}\to f_{0}f_{0} ))\nonumber \\
&V^{(2+,1)}_{\alpha\beta}&=X^{(2)}_{\alpha\beta}(k^\perp)  \\
&2^{++}&\to 2^{++}_{0,+}(0 (2^{++}_{0+}\to f_{0}f_{2} ))\nonumber \\
&V^{(2+,2)}_{\alpha\beta}&=X^{(2)}_{\alpha\beta}(k_{34}^\perp)  \\
&2^{++}&\to 2^{++}_{1,+}(2 (2^{++}_{0+}\to f_{0}f_{2} ))\nonumber \\
&V^{(2+,3)}_{\alpha\beta}&=X^{(2)}_{\alpha\chi}(k^\perp)O_{\beta\chi}^{\mu\nu}X^{(2)}_{\mu\nu}(k_{34}^\perp)  \\
&2^{++}&\to 2^{++}_{0,+}(2 (0^{++}_{2+}\to f_{2}f_{2} ))\nonumber \\
&V^{(2+,4)}_{\alpha\beta}&=X^{(2)}_{\alpha\beta}(k^\perp)X^{(2)}_{\eta\zeta}(k_{12}^\perp) X^{(2)}_{\eta\zeta}(k_{34}^\perp) \\
&2^{++}&\to 2^{++}_{0,+}(0 (2^{++}_{1+}\to f_{2}f_{2} ))\nonumber \\
&V^{(2+,5)}_{\alpha\beta}&=X^{(2)}_{\alpha\zeta}(k_{12}^\perp) X^{(2)}_{\beta\zeta}(k_{34}^\perp) \\
&2^{++}&\to 2^{++}_{1,+}(2 (2^{++}_{1+}\to f_{2}f_{2} ))\nonumber \\
&V^{(2+,6)}_{\alpha\beta}&=X^{(2)}_{\alpha\chi}(k^\perp)O^{\mu\nu}_{\chi
\beta}X^{(2)}_{\mu\zeta}(k_{12}^\perp) X^{(2)}_{\nu\zeta}(k_{34}^\perp) \\
&4^{++}&\to 4^{++}_{0,+}(2 (2^{++}_{0+}\to f_{0}f_{2} ))\nonumber \\
&V^{(4+,1)}_{\alpha\beta\mu\nu}&=X^{(2)}_{\alpha\beta}(k^\perp)O_{\mu\nu}^{\eta\zeta}X^{(2)}_{\eta\zeta}(k_{34}^\perp)  \\
&4^{++}&\to 4^{++}_{0,+}(2 (2^{++}_{1+}\to f_{2}f_{2} ))\nonumber \\
&V^{(4+,2)}_{\alpha\beta\mu\nu}&=X^{(2)}_{\alpha\beta}(k^\perp)O_{\mu\nu}^{\eta\zeta}X^{(2)}_{\eta\chi}(k_{12}^\perp) X^{(2)}_{\chi\zeta}(k_{34}^\perp) \\
&4^{++}&\to 4^{++}_{0,+}(0 (4^{++}_{0+}\to f_{2}f_{2} ))\nonumber \\
&V^{(4+,3)}_{\alpha\beta\mu\nu}&=X^{(2)}_{\alpha\beta}(k_{12}^\perp) X^{(2)}_{\mu\nu}(k_{34}^\perp) \\
&4^{++}&\to 4^{++}_{1,+}(2 (4^{++}_{0+}\to f_{2}f_{2} ))\nonumber \\
&V^{(4+,4)}_{\alpha\beta\mu\nu}&=X^{(2)}_{\alpha\eta}(k^\perp)O_{\eta\beta\mu\nu}^{\zeta\chi\delta\gamma}X^{(2)}_{\zeta\chi}(k_{12}^\perp) X^{(2)}_{\delta\gamma}(k_{34}^\perp)~~~
\end{eqnarray}
Here we provide the amplitudes with $L_4\le 2$  and resonances decaying into two pions up to $J=2$. The unnatural states can only decay into two resonances with $J\ge 2$. In the case of two tensor states, we get the following:
\begin{eqnarray}
&0^{-+}&\to 0^{-+}_{1,+}(1 (1^{++}_{1-}\to f_{2}f_{2} ))\nonumber \\
&V^{(0-,1)}&\!\!=X^{(1)}_{\chi}(k^\perp)\varepsilon_{\chi\mu\nu P}X^{(2)}_{\mu\zeta}(k_{12}^\perp) X^{(2)}_{\nu\zeta}(k_{34}^\perp) \\
&1^{++}&\to 1^{++}_{0,+}(0(1^{++}_{1-}\to f_{2}f_{2} ))\nonumber \\
&V^{(1+,1)}_\alpha&\!\!=\varepsilon_{\alpha\mu\nu P}X^{(2)}_{\mu\zeta}(k_{12}^\perp) X^{(2)}_{\nu\zeta}(k_{34}^\perp) \\
&1^{++}&\to 1^{++}_{1,+}(2(1^{++}_{1-}\to f_{2}f_{2} ))\nonumber \\
&V^{(1+,2)}_\alpha&\!\!=X^{(2)}_{\alpha\beta}(k^\perp)\varepsilon_{\beta\mu\nu P}X^{(2)}_{\mu\zeta}(k_{12}^\perp) X^{(2)}_{\nu\zeta}(k_{34}^\perp) 
\end{eqnarray}
For the cascade decays we provide examples of amplitudes or cascade decays for the orbital momentum $L_4,L_3\le2$:  
\begin{eqnarray}
&J^{PC} &\to J^{PC}_{n,\, \beta} (L_4 (\pi{J_3}_{m \alpha}^{PC} \rightarrow L_3 (\pi f_{J_{12}})) \nonumber \\ 
&0^{+ +}& \rightarrow 0_{0, +}^{+ +}\left( 0\left(\pi 0^{- +}_{0+} \rightarrow 0\left(\pi f_{0} \right) \right)\right) \nonumber\\
&V^{(0+,5)}& \!\!= 1 \\
&0^{+ +} &\rightarrow 0_{1, +}^{+ +}\left( 1\left(\pi 1^{+ +}_{0+} \rightarrow 1\left( \pi f_{0} \right) \right) \right) \nonumber\\
&V^{(0+,6)} &\!\!= X_{\nu}^{(1)}\left( k^{\perp}_4 \right)X_{\nu}^{(1)}\left( k_{3}^{\perp} \right)\\
&0^{+ +}& \rightarrow 0_{2, +}^{+ +}\left(2\left( \pi2^{- +}_{0+} \rightarrow 2\left(\pi f_{0} \right) \right) \right) \nonumber\\
&V^{(0+,7)}& \!\!= X_{\nu\mu}^{(2)}(k^{\perp}_4)X_{\nu\mu}^{(2)}\left( k_{3}^{\perp} \right)\\
&0^{+ +} &\rightarrow 0_{2, +}^{+ +}\left( 2\left(\pi 2^{- +}_{0+} \rightarrow 0\left(\pi f_{2} \right) \right) \right) \nonumber\\
&V^{(0+,8)}& \!\!= X_{\alpha\beta}^{(2)}(k^{\perp}_4)O^{\nu\mu}_{\alpha\beta}(P_3) X_{\nu\mu}^{(2)}\left( k_{12}^{\perp} \right)\\
&0^{+ +}& \rightarrow 0_{1, +}^{+ +}\left( 1\left( \pi1^{+ +}_{1+} \rightarrow 1\left( \pi f_{2} \right) \right) \right) \nonumber \\
&V^{(0+,9)}& \!\!= X_\mu^{(1)}\left( k^{\perp}_4 \right) 
X_{\nu}^{(1)}\left( k_{3}^{\perp} \right)O_{\mu\nu}^{\xi\zeta}(P_3)\times\nonumber \\
&&\quad X_{\xi\zeta}^{(2)}\left( k_{12}^{\perp} \right) \\
&0^{+ +}& \rightarrow 0_{0, +}^{+ +}\left(  0\left( \pi0^{-+}_{2+} \rightarrow 2\left( \pi f_{2} \right) \right) \right)  \nonumber\\
&V^{(0+,10)}& \!\!= X_{\nu_{1}\nu_{2}}^{(2)}\left( k_{3}^{\perp} \right)X_{\nu_{1}\nu_{2}}^{(2)}\left( k_{12}^{\perp} \right)\\
&0^{+ +}& \rightarrow 0_{2, +}^{+ +}\left( 2\left(\pi 2^{- +}_{1+} \rightarrow 2\left(\pi f_{2} \right) \right) \right) \nonumber\\
&V^{(0+,11)}& \!\!= X_{\alpha\beta}^{(2)}( k_{4}^\perp)O_{\alpha\beta}^{\eta\zeta}(P_3)X_{\eta\mu}^{(2)}\left( k_{3}^{\perp} \right)\nonumber \\&&~~~O_{\xi\nu}^{\mu\zeta}(P_3)X^{(2)}_{\xi\nu}\left( k_{12}^{\perp} \right)~~~~~\\ \nonumber
&2^{+ +} &\rightarrow 2_{0, +}^{+ +}\left( 2\left(\pi 0^{- +}_{0+} \rightarrow 0\left(\pi f_{0} \right) \right) \right) \\ 
&V^{(2 + ,7)}_{\alpha\beta}& \!\!= X_{\mu_{1}\mu_{2}}^{(2)}\left( k_{4}^{\perp} \right) \\ \nonumber
&2^{+ +} &\rightarrow 2_{0, +}^{+ +}\left( 1\left(\pi 1^{+ +}_{0+} \rightarrow 1\left(\pi  f_{0} \right) \right) \right) \\
&V^{(2 + ,8)}_{\alpha\beta} &\!\!= X_{\alpha^{}}^{(1)}\left( k_{4}^{\perp} \right)X_{\beta^{}}^{(1)}\left( k_{3}^{\perp} \right)
\\ 
\nonumber
&2^{+ +}& \rightarrow 2_{0, +}^{+ +}\left( 0\left(\pi 2^{- +}_{0+} \rightarrow 2\left(\pi  f_{0} \right) \right) \right) \\
&V^{(2 + ,9)}_{\alpha\beta}& \!\!= X_{\alpha\beta}^{(2)}\left( k_{3}^{\perp} \right)\\ \nonumber
&2^{+ +} &\rightarrow 2_{1, +}^{+ +}\left( 2\left(\pi 2^{- +}_{0+} \rightarrow 2\left(\pi  f_{0} \right) \right) \right) \\
&V^{(2 + ,10)}_{\alpha\beta}& \!\!= X_{\alpha\nu}^{(2)}\left( k_{4}^{\perp} \right)X_{\nu\beta}^{(2)}\left( k_{3}^{\perp} \right)\\ \nonumber
&2^{+ +}& \rightarrow 2_{0, +}^{+ +}\left( 0\left(\pi 2^{- +}_{0+} \rightarrow 0\left(\pi  f_{2} \right) \right) \right) \\
&V^{(2 + ,11)}_{\alpha\beta}& \!\!= O^{\mu\nu}_{\alpha\beta}(P_3) X_{\mu\nu}^{(2)}\left( k_{12}^{\perp} \right)
\end{eqnarray}
\begin{eqnarray}
\nonumber
&2^{+ +}& \rightarrow 2_{1, +}^{+ +}\left( 2\left(\pi 2^{- +}_{0+} \rightarrow 0\left(\pi  f_{2} \right) \right) \right) \\
&V^{(2 + ,12)}_{\alpha\beta}&\!\!= X_{\alpha\nu}^{(2)}\left( k_{4}^{\perp} \right)O^{\nu\beta}_{\xi\zeta}(P_3)X_{\xi\zeta}^{(2)}\left( k_{12}^{\perp} \right)
\\ 
\nonumber
&2^{+ +} &\rightarrow 2_{0, +}^{+ +}\left( 1\left(\pi 1^{+ +}_{1+} \rightarrow 1\left(\pi  f_{2} \right) \right) \right) 
\\
&V^{(2 + ,13)}_{\alpha\beta}& \!\!= 
X_{\alpha^{}}^{(1)}\left( k_{4}^{\perp} \right)
X_{\nu_{1}}^{(1)}\left( k_{3}^{\perp} \right)\times
 \nonumber \\
&&\quad O_{\beta\nu_{1}}^{\rho_{1}\rho_{2}}(P_3)
X_{\rho_{1}\rho_{2}^{}}^{(2)}\left( k_{12}^{\perp} \right)\\ \nonumber
&2^{+ +} &\rightarrow 2_{0, -}^{+ +}\left( 1\left(\pi 2^{+ +}_{0-} \rightarrow 1\left(\pi  f_{2} \right) \right) \right) \\ \nonumber
&V^{(2 + ,14)}_{\alpha\beta}& = 
\varepsilon_{\nu_{1}\rho_{2}\alpha^{}P}
X_{\nu_{1}}^{(1)}\left( k_{4}^{\perp}\right)
O_{\rho_{3}\rho_{4}}^{\rho_{2}\beta^{}}(P_3)\times
\\
&&\quad \varepsilon_{\nu_{2}\nu_{3}\rho_{3}P_3}X_{\nu_{2}}^{(1)}\left( k_{3}^{\perp} \right)
X_{\nu_{3}\rho_{4}}^{(2)}\left( k_{12}^{\perp} \right)
\\ \nonumber
&2^{+ +} &\rightarrow 2_{0, +}^{+ +}\left( 2\left(\pi 0^{- +}_{2+} \rightarrow 2\left(\pi f_{2} \right) \right) \right) \\
&V^{(2 + ,15)}_{\alpha\beta}& = X_{\alpha\beta}^{(2)}\left( k_{4}^{\perp} \right)X_{\nu_{1}\nu_{2}}^{(2)}\left( k_{3}^{\perp} \right)X_{\nu_{1}\nu_{2}}^{(2)}\left( k_{12}^{\perp} \right)\\ \nonumber
&2^{+ +} &\rightarrow 2_{0, +}^{+ +}\left( 0\left(\pi 2^{- +}_{1+} \rightarrow 2\left(\pi f_{2} \right) \right) \right) \\
&V^{(2 + ,16)}_{\alpha\beta}& = 
O_{\alpha\beta}^{\rho_{1}^{}\rho_{2}^{}}(P_3)
X_{\rho_{1}^{}\nu_{1}}^{(2)}\left( k_{3}^{\perp} \right)
O_{\mu\nu}^{\nu_{1}\rho_{2}}(P_3)\times\nonumber \\
&&\quad  X_{\mu\nu}^{(2)}\left( k_{12}^{\perp} \right)
\\ \nonumber
&2^{+ +} &\rightarrow 2_{1, +}^{+ +}\left( 2\left(\pi 2^{- +}_{1+} \rightarrow 2\left(\pi f_{2} \right) \right) \right) \\
&V^{(2 + ,17)}_{\alpha\beta}&= 
X_{\rho_{1}\alpha}^{(2)}\left( k_{4}^{\perp} \right)
O_{\rho_{3}\rho_{2}}^{\beta\rho_{1}}(P_3)\times
\nonumber \\
&&\quad X_{\rho_{3}\nu_{1}}^{(2)}\left( k_{3}^{\perp} \right)
O^{\rho_4 \rho_5}_{\nu_{1}\rho_{2}} (P_3) X_{\rho_4 \rho_5}^{(2)}\left( k_{12}^{\perp} \right)
\end{eqnarray}

\begin{eqnarray}
&2^{- +}& \rightarrow 2_{0, -}^{- +}\left( 2\left(\pi 1^{+ +}_{0+} \rightarrow 1\left(\pi f_{0} \right) \right) \right) \nonumber \\
 &V^{(2 - ,1)}_{\alpha\beta} &= 
\varepsilon_{\nu_{1}\nu_{2}\alpha^{}P}
X_{\nu_{1}\beta^{}}^{(2)}\left( k_{4}^\perp \right)
X_{\nu_{2}}^{(1)}\left( k_{3}^\perp \right)\\ \nonumber
&2^{- +} &\rightarrow 2_{0, -}^{- +}\left( 1\left(\pi 2^{- +}_{0+} \rightarrow 2\left(\pi f_{0} \right) \right) \right) \\
&V^{(2 - ,2)}_{\alpha\beta}& = 
\varepsilon_{\nu_{1}\nu_{2}\alpha^{}P}X_{\nu_{1}}^{(1)}\left( k_{4}^\perp \right)
X_{\nu_{2}\beta^{}}^{(2)}\left( k_{3}^\perp \right)
\\ \nonumber
&2^{- +}& \rightarrow 2_{0, -}^{- +}\left( 1\left(\pi 2^{- +}_{0+} \rightarrow 0\left(\pi f_{2} \right) \right) \right) \\
&V^{(2 - ,3)}_{\alpha\beta}& = 
\varepsilon_{\nu_{1}\nu_{2}\alpha^{}P}
X_{\nu_{1}}^{(1)}\left( k_{4}^\perp \right) O_{\nu_2 \beta}^{\rho_1 \rho_2}(P_3) \times \nonumber \\
&&X_{\rho_1 \rho_2}^{(2)}\left( k_{12}^\perp \right)
\\ \nonumber
&2^{- +} &\rightarrow 2_{0, -}^{- +}\left( 2\left(\pi 1^{+ +}_{1+} \rightarrow 1\left(\pi f_{2} \right) \right) \right) \\
&V^{(2 - ,4)}_{\alpha\beta} &= 
\varepsilon_{\nu_{1}\nu_{2}\alpha^{}P}
X_{\nu_{1}\beta^{}}^{(2)}\left( k_{4}^\perp \right)
X_{\nu_{3}}^{(1)}\left( k_{3}^\perp \right)\times
\nonumber \\ 
&&\quad O_{\nu_{3}\nu_{2}}^{\rho_{1}^{}\rho_{2}^{}}(P_3)
X_{\rho_{1}\rho_{2}}^{(2)}\left( k_{12}^\perp \right)
\\ \nonumber
&2^{- +}& \rightarrow 2_{0, +}^{- +}\left( 0\left(\pi 2^{+ +}_{0-} \rightarrow 1\left(\pi f_{2} \right) \right) \right) \\
&V^{(2 - ,5)}_{\alpha\beta}& = 
\varepsilon_{\nu_{1}\nu_{2}\alpha^{}P}
X_{\nu_{1}}^{(1)}\left( k_{3}^\perp \right)\times
\nonumber \\
&&\quad O_{\nu_{2}\beta}^{\rho_{1}^{}\rho_{2}^{}}(P_3)
X_{\rho_{1}\rho_{2}^{}}^{(2)}\left( k_{12}^\perp \right)
\\ \nonumber
&2^{- +}& \rightarrow 2_{1, +}^{- +}\left( 2\left(\pi 2^{+ +}_{0-} \rightarrow 1\left(\pi f_{2} \right) \right) \right) \\
&V^{(2 - ,6)}_{\alpha\beta}& = 
X_{\rho_{2}\alpha^{}}^{(2)}\left( k_{4}^\perp \right)
O_{\rho_{3}\rho_{1}}^{\rho_{2}\beta^{}}(P_3)\times
\nonumber \\ 
&&\quad \varepsilon_{\nu_{1}\nu_{2}\rho_{3}P_3}
X_{\nu_{1}}^{(1)}\left( k_{3}^\perp \right)
X_{\nu_{2}\rho_{1}}^{(2)}\left( k_{12}^\perp \right)
\\ \nonumber
&2^{- +}& \rightarrow 2_{0, -}^{- +}\left( 1\left(\pi 2^{- +}_{1+} \rightarrow 2\left(\pi f_{2} \right) \right) \right) \\
&V^{(2 - ,7)}_{\alpha\beta} &= 
X^{(1)}_{\nu_{1}}(k_4^{\perp})
\varepsilon_{\nu_{1}\rho_{2}\alpha^{}P_3} 
O_{\rho_{3}\rho_{1}}^{\rho_{2}\beta^{}}(P_3)\times
\nonumber \\
&&\quad X_{\nu_{2}\rho_{3}}^{(2)}\left( k_{3}^\perp \right)
 O^{\rho_4 \rho_5}_{\nu_{2}\rho_{1}} (P_3) X_{\rho_4 \rho_5}^{(2)}\left( k_{12}^\perp \right)
\nonumber \\
&0^{- +}& \rightarrow 0_{2, +}^{- +}\left( 2\left(\pi 2^{+ +}_{0-} \rightarrow 1\left(\pi f_{2} \right) \right) \right)
\nonumber \\
\end{eqnarray}
\begin{eqnarray}
&V^{(0 - ,2)} &= 
X_{\rho_{1}^{}\rho_{2}^{}}^{(2)}\left( k_{4}^\perp \right)
O_{\rho_{3}^{}\rho_{4}^{}}^{\rho_{1}^{}\rho_{2}^{}}(P_3) \varepsilon_{\nu_{1}\nu_{2}\rho_{3}^{}P_3}\times
\nonumber \\
&&\quad 
X_{\nu_{1}}^{(1)}\left( k_{3}^\perp \right) O^{\rho_5 \rho_6}_{\nu_{2}\rho_{4}} (P_3)
X_{\rho_5 \rho_6}^{(2)}\left( k_{12}^\perp \right)
\nonumber \\
&1^{+ +}& \rightarrow 1_{0, -}^{+ +}\left( 1\left(\pi 1^{+ +}_{0+} \rightarrow 1\left(\pi f_{0} \right) \right) \right)  \nonumber \\
&V^{(1 + ,3)}_{\alpha}& = \varepsilon_{\alpha\nu_{1}\nu_{2}P}X_{\nu_{1}}^{(1)}\left( k_{4}^\perp \right)X_{\nu_{2}}^{(1)}\left( k_{3}^\perp \right) \\ \nonumber
&1^{+ +} &\rightarrow 1_{1, -}^{+ +}\left( 2\left(\pi 2^{- +}_{0+} \rightarrow 2\left(\pi f_{0} \right) \right) \right) \\
&V^{(1 + ,4)}_{\alpha}& = \varepsilon_{\alpha\nu_{1}\nu_{2}P}X_{\nu_{1}\nu_{3}}^{(2)}\left( k_{4}^\perp \right)X_{\nu_{2}\nu_{3}}^{(2)}\left( k_{3}^\perp \right)\\ \nonumber
&1^{+ +}& \rightarrow 1_{1, -}^{+ +}\left( 2\left(\pi 2^{- +}_{0+} \rightarrow 0\left(\pi f_{2} \right) \right) \right) \\
&V^{(1 + ,5)}_{\alpha}& = \varepsilon_{\alpha\nu_{1}\nu_{2}P}X_{\nu_{1}\nu_{3}}^{(2)}\left( k_{4}^\perp \right)O^{\rho_1 \rho_2}_{\nu_{2}\nu_{3}}(P_3)\times\nonumber \\&&X_{\rho_1 \rho_2}^{(2)}\left( k_{12}^\perp \right)\\ \nonumber
&1^{+ +} &\rightarrow 1_{0, -}^{+ +}\left( 1\left(\pi 1^{+ +}_{1+} \rightarrow 1\left(\pi f_{2} \right) \right) \right) \\
&V^{(1 + ,6)}_{\alpha}& = 
\varepsilon_{\alpha\nu_{1}\nu_{2}P}
X_{\nu_{1}}^{(1)}\left( k_{4}^\perp \right)X_{\nu_{3}}^{(1)}\left( k_{3}^\perp \right)\times
\nonumber \\
&&\quad O_{\nu_{3}\nu_{2}}^{\rho_{1}\rho_{2}}(P_3)
X_{\rho_{1}\rho_{2}}\left( k_{12}^\perp \right)
\\ \nonumber
&1^{+ +} &\rightarrow 1_{1, +}^{+ +}\left( 1\left(\pi 2^{+ +}_{0-} \rightarrow 1\left(\pi f_{2} \right) \right) \right) \\ \nonumber
&V^{(1 + ,7)}_{\alpha}& = 
X_{\rho_{1}^{}}^{(1)}\left( k_{4}^\perp \right)
O_{\alpha\rho_{1}^{}}^{\rho_{3}^{}\rho_{2}^{}}(P_3)\varepsilon_{\nu_{1}\nu_{2}\rho_{2}^{}P_3} \times
\\ 
&&\quad 
X_{\nu_{1}}^{(1)}\left( k_{3}^\perp \right)
O^{\rho_4 \rho_5}_{\nu_{2}\rho_{3}} (P_3) X_{\rho_4 \rho_5}^{(2)}\left( k_{12}^\perp \right)
\\ \nonumber
&1^{+ +}& \rightarrow 1_{1, -}^{+ +}\left( 2\left(\pi 2^{- +}_{1+} \rightarrow 2\left(\pi f_{2} \right) \right) \right) \\
&V^{(1 + ,8)}_{\alpha}& = 
\varepsilon_{\alpha\nu_{1}\rho_{1}^{}P}
X_{\nu_{1}\rho_{2}^{}}^{(2)}\left( k_{4}^\perp \right)
O_{\rho_{3}^{}\rho_{4}^{}}^{\rho_{1}^{}\rho_{2}^{}}(P_3)\times
\nonumber \\ 
&&\quad X_{\nu_{2}\rho_{3}^{}}^{(2)}\left( k_{3}^\perp \right)
O^{\rho_5 \rho_6}_{\nu_{2}\rho_{4}} (P_3) X_{\nu_{2}\rho_{4}}^{(2)}\left( k_{12}^\perp \right)
\end{eqnarray}
\begin{eqnarray}
&4^{+ +} &\rightarrow 4_{0, +}^{+ +}\left( 2\left(\pi 2^{- +}_{0+} \rightarrow 2\left(\pi f_{0} \right) \right) \right) \nonumber \\
&V^{(4 + ,5)}_{\alpha\beta\mu\nu}& = 
X_{\alpha^{}\beta^{}}^{(2)}\left( k_{4}^\perp \right)
X_{\mu^{}\nu^{}}^{(2)}\left( k_{3}^\perp \right) \\
&4^{+ +} &\rightarrow 4_{0, +}^{+ +}\left( 2\left(\pi 2^{- +}_{0+} \rightarrow 0\left(\pi f_{2} \right) \right) \right) \nonumber \\
&V^{(4 + ,6)}_{\alpha\beta\mu\nu} &= 
X_{\alpha^{}\beta^{}}^{(2)}\left( k_{4}^\perp \right) O^{\rho_1 \rho_2}_{\mu^{}\nu^{}} (P_3)
X_{\rho_1 \rho_2}^{(2)}\left( k_{12}^\perp \right)
\\
&4^{+ +}& \rightarrow 4_{0, +}^{+ +}\left( 2\left(\pi 2^{- +}_{1+} \rightarrow 2\left(\pi f_{2} \right) \right) \right) 
\nonumber\\
&V^{(4 + ,7)}_{\alpha\beta\mu\nu} &= 
X_{\alpha^{}\beta^{}}^{(2)}\left( k_{4}^\perp \right)
O^{\rho_{3}\rho_{4}}_{\mu\nu}(P_3)\times
\nonumber \\
&&\quad X_{\nu_{1}\rho_{3}}^{(2)}\left( k_{3}^\perp \right)
O^{\rho_1 \rho_2}_{\nu_{1}\rho_{4}} (P_3)
X_{\rho_1 \rho_2}^{(2)}\left( k_{12}^\perp \right)
\end{eqnarray}

\subsection{The transition amplitudes for the central production due to pomeron-pomeron ($f_0f_0$) collision.}
In the pomeron-pomeron central collision (or in $f_0f_0$ collision) only isoscalar states with even spin, positive  $P,C$-parity are produced. The production vertices are described by the orbital momentum tensors. Therefore, we obtain the following expressions for the partial-wave amplitudes:
\begin{eqnarray}
    A^{(0+,i)}=&V^{(0+,i)} & i=1\!-\!11, \nonumber \\
    A^{(2+,i)}=&X^{(2)}_{\mu\nu }(q^\perp)O^{\mu\nu }_{\alpha\beta}V_{\alpha\beta}^{(2+,i)} & i=1\!-\!17, \nonumber \\
    A^{(4+,i)}=&X^{(4)}_{\mu\nu\chi\xi}(q^\perp)O^{\alpha\beta\eta\zeta}_{\mu\nu\chi\xi}V^{(4+,i)}_{\alpha\beta\eta\zeta} & i=1\!-\!7 ~~~~~~
\end{eqnarray}
\subsection{The radiative $J/\Psi$ decay into $4\pi^0$ final state. }
In the case of the radiative $J/\psi$ it is usually assumed that after emission of the photon the $c\bar c$ system is annihilated to a resonance which decays into hadron final states. From this point of view, all states that are allowed in the quark model can be produced. These states have isospin $I=0$ and charged parity $C=+1$. Let us give the list of the amplitudes up to $J=4$:      
\begin{eqnarray}
    A^{(0+,i)}=&S^0V^{(0+,i)} & i=1\!-\!11,  \nonumber \\
    A_1^{(2+,i)}=&S^0X^{(2)}_{\mu\nu }(q^\perp)O^{\mu\nu }_{\alpha\beta}V_{\alpha\beta}^{(2+,i)} & i=1\!-\!17, \nonumber \\
    A_2^{(2+,i)}=&\varepsilon_{\mu\eta\xi P}S^1_\eta X^{(2)}_{\xi\nu\
    }(q^\perp)O^{\mu\nu }_{\alpha\beta}V_{\alpha\beta}^{(2+,i)} & i=1\!-\!17, \nonumber \\
    A_3^{(2+,i)}=&S^2_{\mu\nu }O^{\mu\nu }_{\alpha\beta}V_{\alpha\beta}^{(2+,i)} & i=1\!-\!17, \nonumber \\
    A_1^{(4+,i)}=&S^0X^{(4)}_{\mu\nu\alpha\beta}(q^\perp)O^{\mu\nu\alpha\beta}_{\eta\chi\zeta\varrho}V_{\eta\chi\zeta\varrho}^{(4+,i)} & i=1\!-\!7, \nonumber \\
    A_2^{(4+,i)}=&\varepsilon_{\mu\eta\xi P}S^1_\eta X^{(4)}_{\xi\nu\alpha\beta
    }(q^\perp)O^{\mu\nu\alpha\beta}_{\chi\zeta\varrho\delta}V_{\chi\zeta\varrho\delta}^{(4+,i)} & i=1\!-\!7, \nonumber \\
    A_3^{(4+,i)}=&S^2_{\mu\nu}X^{(2)}_{\alpha\beta}(q^\perp)O^{\mu\nu\alpha\beta}_{\eta\chi\zeta\varrho}V_{\eta\chi\zeta\varrho}^{(4+,i)} & i=1\!-\!7, \nonumber \\
    A^{(0-,i)}= &S^1_\eta X^{(1)}_{\eta}(q^\perp) V^{(0-,i)}_{} & i=1\!-\!2 \nonumber \\
    A_1^{(1+,i)}= &S^1_{\mu}V^{(1+,i)}_{\mu} & i=1\!-\!8  \nonumber \\
    A_2^{(1+,i)}= &S^1_{\eta} X^{(2}_{\eta\mu}(q^\perp)V^{(1+,i)}_{\mu} & i=1\!-\!8  \nonumber \\
    A_1^{(2-,i)}= &S^1_{\mu} X^{(1)}_{\nu}(q^\perp)O_{\mu\nu}^{\alpha\beta}  
    V^{(2-,i)}_{\alpha\beta} & i=1\!-\!7  \nonumber \\
    A_2^{(2-,i)}= &S^1_{\eta} X^{(3)}_{\eta\mu\nu}(q^\perp)O_{\mu\nu}^{\alpha\beta}  
    V^{(2-,i)}_{\alpha\beta} & i=1\!-\!7  \nonumber \\
    A_3^{(2-,i)}= & \varepsilon_{\mu\eta \xi P}S^2_{\eta\nu} X^{(1)}_{\xi}(q^\perp)    
    O^{\alpha\beta}_{\mu\nu}   V^{(2-,i)}_{\alpha\beta} & i=1\!-\!7  \nonumber
    \\
    &&
\end{eqnarray}

\bibliography{refer}
\end{document}